\documentclass[pra,preprint,superscriptaddress,floatfix,showpacs, 11pt]{revtex4-1}
\usepackage[T1]{fontenc}
\usepackage{geometry}
\usepackage{amsmath,amsfonts,amssymb}
\usepackage{graphicx}
\usepackage{bm}
\usepackage{xcolor}
\newgeometry{vmargin={25mm}, hmargin={21mm,22mm}}

\begin{document}

\title{Rotation quenches in trapped bosonic systems }
\author{Rhombik Roy}
\email{rroy@campus.haifa.ac.il}
\affiliation{Department of Physics, University of Haifa, Haifa 3498838, Israel}
\affiliation{Haifa Research Center for Theoretical Physics and Astrophysics, University of Haifa,
Haifa 3498838, Israel}
\author{Sunayana Dutta}
\affiliation{Department of Physics, University of Haifa, Haifa 3498838, Israel}
\affiliation{Haifa Research Center for Theoretical Physics and Astrophysics, University of Haifa,
Haifa 3498838, Israel}
\author{Ofir E. Alon}
\affiliation{Department of Physics, University of Haifa, Haifa 3498838, Israel}
\affiliation{Haifa Research Center for Theoretical Physics and Astrophysics, University of Haifa,
Haifa 3498838, Israel}

\date{\today}

\begin{abstract}
 
The ground state properties of strongly rotating bosons confined in an asymmetric anharmonic potential exhibit a split density distribution. However, the out-of-equilibrium dynamics of this split structure remain largely unexplored. Given that rotation is responsible for the breakup of the bosonic cloud, we investigate the out-of-equilibrium dynamics by abruptly changing the rotation frequency. Our study offers insights into the dynamics of trapped Bose-Einstein condensates in both symmetric and asymmetric anharmonic potentials under different rotation quench scenarios. In a rotationally symmetric trap, angular momentum is a good quantum number. Therefore, a rotation quench has no impact on the density distribution. In contrast, the absence of angular momentum conservation in asymmetric traps results in more complex dynamics. This allows rotation quenches to either inject into or extract angular momentum from the system. We observe and analyze these intricate dynamics both for the mean-field condensed and the many-body fragmented systems. The dynamical evolution of the condensed system and the fragmented system exhibits similarities in several observables during small rotation quenches. However, these similarities diverge notably for larger quenches. Additionally, we investigate the formation and the impact of the vortices on the angular momentum dynamics of the evolving split density. All in all, our findings offer valuable insights into the dynamics of trapped interacting bosons under different rotation quenches. 
\end{abstract}

\maketitle

\section{Introduction}\label{intro} 
Rotation has remarkable effects, influencing phenomena from vast astronomical bodies like galaxies and massive rotating stars~\cite{rotation_ingalaxy1,rotation_ingalaxy2} to the intricate quantum domain of atomic nuclei~\cite{rotation_nuclei}. Ultracold bosonic gases under rotation offer a promising avenue for exploring various properties inherent in correlated quantum systems~\cite{rotation_c1,rotation_c2,rotation_c3,rotation_c4,rotation_c5,gp2}. These include phenomena like quantized vortices~\cite{rotation_quantized_vortex}, vortex nucleation~\cite{rotation_vortex_nucleation}, the emergence of quantum fluctuations~\cite{rotation_quantum_fluctuation1,rotation_quantum_fluctuation2}, and the manifestation of the fractional quantum Hall effect in weakly interacting quantum systems~\cite{rotation_quantum_hall_effect}. Numerous studies have investigated the properties of rotating quantum gases. For instance, experiments employing anharmonic potentials have revealed the emergence of superfluid flow in atomic physics~\cite{rotation_superfluif_flow}. Ring-shaped traps have proven especially valuable to induce and maintain persistent currents in Bose-Einstein condensates (BECs)~\cite{rotation_superfluid_flow2,rotation_superfluid_flow3,rotation_superfluid_flow4}. 
On the theoretical front, investigations have explored the dynamics of rotating dipolar condensates, encompassing both three-dimensional configurations~\cite{rotation_3d} as well as in a rotating, asymmetric, pancake-like quartic-quadratic potential~\cite{rotation_gammal}.

Studies have shown that rotation influences the breaking up of the ground state density of weakly interacting bosons~\cite{rotation_gammal,beinke_rotation2,sunayana_scirep}. But the process of how angular momentum is acquired and ejected from the system after a quench remains largely unexplored. As the density of the system splits due to the influence of rotation, we are keen to explore how an abrupt change in the rotation frequency drives the dynamical evolution of such systems. The dynamics in the rotating systems are profoundly interesting as they exhibit intricate interplay between rotational effects and many-body correlations. Indeed, numerous studies have utilized many-body treatments to explore fragmentation and correlation in BECs both in non-rotating frames~\cite{mctdhb_fragmentation1,rhombik_epjd,mctdhb_fragmentation3,rhombik_quantumreports,mctdhb_fragmentation4} and in rotating frames~\cite{many-body-rotation1,many-body-rotation2,many-body-rotation3,many-body-rotation4,many-body-rotation5,rotation_budha}. However, the investigation of many-body features in the rotating frame has received comparatively less attention. 
Typically, the investigation of rotating bosons often relies on the Gross-Pitaevskii (GP) method, which considers only a single eigenvalue of the one-body reduced density matrix~\cite{gp1,gp2,gp3}. According to Penrose and Onsager, interacting bosons are fully condensed if they exhibit a single macroscopic eigenvalue of the reduced one-particle density matrix~\cite{penrose-onsager}. Hence, the GP method treats bosons as condensed and overlooks the presence of correlations and fragmentation in the system. Consequently, the study of rotating bosons necessitates a dedicated many-body treatment to comprehensively understand the intricate dynamics, correlations, and fragmentation phenomena.
 
In this study, we have explored the dynamics after a rotation quench in trapped bosons in two-dimensional anharmonic potentials. In a harmonic trapping potential, when the rotation frequency is comparable to the trapping frequency, the centrifugal force acting on the bosons due to rotation can surpass the trapping force, allowing the bosons to escape the potential~\cite{angular_limit}. To mitigate this, we have used anharmonic traps throughout our calculations~\cite{asy_trap_1,asy_trap_2}. Rotation in trapped BECs can be controlled using rotating magnetic traps or laser stirring techniques, enabling the experimental exploration of rotation quench dynamics in these systems~\cite{PhysRevLett.97.240402}. In our numerical calculations, the initial state is prepared with an initial rotation frequency featuring split density distribution. The rotation frequency is then abruptly lowered to observe the subsequent dynamics. Broadly, two scenarios are examined: bosons trapped in a symmetric potential and bosons trapped in asymmetric potentials. Due to angular momentum conservation, all observables exhibit monotonic behavior during the quench dynamics in the symmetric trap. However, when the rotational symmetry of the trap is broken and the trap is elongated along the x-direction, intriguing effects emerge following the rotation quench. These effects include the formation of vortices in the density distribution, oscillations in the angular momentum, its connection to the density dynamics, and, in some cases, the buildup or loss of coherence. The dynamics of the system in response to the rotation quenches after partially restoring the symmetry with respect to the elongated potential are also examined. We investigate the dynamics for both mean-field condensed and many-body fragmented systems and explore the regimes where the dynamics of the condensed systems significantly deviate from the fragmented systems. Overall, we provide insights into how the system of weakly repulsively interacting bosons trapped in anharmonic trapping potential responds to different rotation quenches.

The multiconfigurational time-dependent Hartree method for bosons (MCTDHB), a bosonic adaptation of the MCTDH family of methods~\cite{rhombik_pre,cpl1990,jcp1992,mctdhb12,mctdhb16,mctdhb17,fischer_Metrology,paolo_cavity,paolo_ol,rhombik_pra,spin_axel,elke,rhombik_jpb,rhombik_acc}, is widely recognized for its ability to deliver highly accurate results~\cite{mctdhb_review}. The many-body MCTDHB method can capture both the fragmentation of bosons and the correlations between them in a self-consistent manner~\cite{MCTDHB1,MCTDHB2}, making it a valuable tool to explore intricate dynamical phenomena in the bosonic systems. In our numerical calculations, we employ the MCTDHB method to solve the many-body Schr\"odinger equation, implemented in the MCTDH-X software~\cite{mctdhb_software1,mctdhb_software2}.

The paper is organized as follows. In Section~\ref{method}, we introduce the setup and necessary theoretical background. We also discuss the fundamental equations used to measure various quantities of interest. Section~\ref{result} presents our numerical results and interpretations for rotation quenches in three different trapping potentials. We investigate one case with a rotationally symmetric anharmonic potential and two cases introducing asymmetry: an elongated trapping potential and a four-fold symmetric trapping potential. Finally, Section~\ref{conclusion} summarizes our findings. 
The supplementary material includes a brief discussion on the MCTDHB method, the characterization of different rotation quench magnitudes, the numerical convergence analysis, and the discussion of the large rotation quench in the four-fold symmetric trap. It also provides full density and vortex dynamics movies calculated using both mean-field and many-body methods.

\section{System setup}\label{method}
In our calculation, we focus on weakly interacting bosons confined in a two-dimensional anharmonic trap potential under rotation. Our study examines the rotation quench dynamics in three types of anharmonic trapping potentials: a rotationally symmetric trap, an elongated trap along the x-direction that breaks the symmetry, and a four-fold symmetric trap that partially ``restores" the symmetry of the second one. The dynamics after a rotational quench are studied by solving the time-dependent Schr\"odinger equation, $\hat{H}\vert \Psi \rangle = i \partial_t \vert \Psi \rangle$. The form of the Hamiltonian governing $N$ interacting bosons in the rotating frame is given by:
\begin{equation} \label{hamiltonian}
    \hat{H}= \sum_{i=1}^{N} \hat{h}_i+ \sum_{i<j} \hat{W}(\mathbf{r}_i-\mathbf{r}_j) -\omega \hat{L}_Z.
\end{equation}
The first term is the single particle Hamiltonian and has the form $\hat{h}= -\frac{1}{2} \frac{\partial^2}{\partial \mathbf{r}^2} + V_{trap} (\mathbf{r}) $. $V_{trap} (\mathbf{r})$ is the external two-dimensional trapping potential; $\mathbf{r} = (x, y)$ denotes the position vector in two spatial dimensions. 
The second term in the Hamiltonian accounts for the interactions between the bosons. In our calculations, we use a Gaussian-shaped repulsive interaction given by $\hat{W}(\mathbf{r}_i-\mathbf{r}_j) = \frac{\lambda_0}{2 \pi \sigma^2} e^{-\frac{(\mathbf{r}_i - \mathbf{r}_j)^2}{2\sigma^2}}$. The interaction width is set to $\sigma=0.25$ to circumvent the regularization problem intrinsic to the zero-range contact potential in two dimensions~\cite{gaussian_interaction1,gaussian_interaction2}. This choice has been employed in recent research works~\cite{sunayana_scirep,anal2020}. The interaction strength is $\lambda_0$ and the mean-field interaction parameter is defined as $\Lambda = \lambda_0 (N-1)$, where $N$ is the number of bosons. 
The third term in the Hamiltonian accounts for the rotation in the system, which encompasses both the rotation frequency $\omega$ and the total angular momentum operator in the z-direction, denoted as $\hat{L}_Z$. We solve the dimensionless Schr\"odinger equation by scaling the Hamiltonian $\hat{H}$ by $\frac{\hbar^2}{mL^2}$, where $L$ is a convenient length scale and $m$ is the mass of the boson. We adopt natural units, where $\hbar = m = 1$. In real systems like cold-atom experiments, measurements follow physical scales and system sizes are typically on the order of micrometers. For example, using $L = 1\mu$m and $m = 1.4431 \times 10^{-25}$ kg for a $^{87}$Rb atom, the time unit $\hbar^2 / (mL^2) = 1.37$ ms. So, $1$ sec = $729.92$ time units and $1$ m = $10^6$ length units.

\subsection{Key measures}
Using the $N$-body wave-function $\Psi (\mathbf{r}_1,\mathbf{r}_2, \dots ,\mathbf{r}_N;t)$, obtained by solving the time-dependent Schr\"odinger equation, we calculate the one-body density evolution, occupations in the natural orbitals, the average angular momentum, and the variance of several observables to analyze the post-quench dynamics.

\subsubsection{Reduced density matrices, occupation in natural orbitals and one-body density}
The reduced one-body density matrix (RDM) can be constructed from the wave-function $\Psi (\mathbf{r}_1,\mathbf{r}_2, \dots ,\mathbf{r}_N)$ as
\begin{equation}\label{rdm}
\begin{split}
    \frac{\rho^{(1)}(\mathbf{r},\mathbf{r}^\prime)}{N}=\int d\mathbf{r}_2 \dots \mathbf{r}_N \Psi^* (\mathbf{r}^\prime,\mathbf{r}_2, \dots ,\mathbf{r}_N) 
    \Psi (\mathbf{r},\mathbf{r}_2, \dots ,\mathbf{r}_N) 
    = \sum_j \frac{n_j}{N} \alpha_j(\mathbf{r}) \alpha_j^* (\mathbf{r}^\prime).
\end{split}
\end{equation}
Here, $\alpha_j (\mathbf{r})$ are the natural orbitals and $n_j$ is the occupation in the respective natural orbitals. The one-body density is determined by taking the diagonal elements of the reduced one-body density matrix, which is expressed as $\rho(\mathbf{r}) = \rho^{(1)}\left(\mathbf{r},\mathbf{r}^\prime = \mathbf{r} \right)$.

\subsubsection{Angular momentum}
Analyzing angular momentum is crucial for understanding rotational dynamics in two dimensions. It has been extensively utilized to investigate various rotating systems~\cite{angular101,angular102}, as well as in the study of hidden vortices~\cite{phantom_angular_axel}, spatially partitioned vortices~\cite{beinke_rotation1}, and fragmentation phenomena~\cite{beinke_rotation1,beinke_rotation2,phantom_angular_axel}. In a two dimensional system in the x-y plane, the average angular momentum is given by $\langle \Psi(\mathbf{r_1},\mathbf{r}_2, \dots ,\mathbf{r}_N) \vert \hat{L}_Z \vert \Psi(\mathbf{r_1},\mathbf{r}_2, \dots ,\mathbf{r}_N) \rangle$, where
\begin{equation}\label{eq.angularmom2}
    \hat{L}_Z = \sum_{j=1}^{N} \hat{l}_{z_j} = \sum_{j=1}^N \frac{1}{i} \left( \hat{x}_j \frac{\partial}{\partial y_j} - \hat{y}_j \frac{\partial}{\partial x_j} \right).
\end{equation}
In rotationally symmetric trap potentials, angular momentum is conserved quantity. However, in an asymmetric trapping potential, this symmetry is broken, allowing the particles to exchange angular momentum during the time evolution. 

\subsubsection{Many-body variances}
To explore the correlations within the system, calculating variance of certain observables is a powerful method~\cite{variance_ofir2015}. For an observable $\hat{A} = \sum_{j=1}^N \hat{a} (\mathbf{r}_j)$, the variance is given by
\begin{equation}\label{eq.variance1}
    \frac{1}{N}\Delta_{\hat{A}}^2 = \frac{1}{N} \left( \langle\hat{A}^2 \rangle - \langle\hat{A} \rangle^2 \right),
\end{equation}
where $\langle \hat{A} \rangle$ depends only on the one-body operator. The second term in Eq.~\ref{eq.variance1} has contribution from the one- and two-body operators
\begin{equation}\label{eq.variance2}
    \hat{A}^2 = \sum_{j=1}^N \hat{a}^2 (\mathbf{r}_j) + 2\sum_{j<k} 2 \hat{a} (\mathbf{r}_j) \hat{a} (\mathbf{r}_k).
\end{equation}
Thus, a small fraction of particles outside the condensate can interact with the entire condensate, allowing variance to effectively quantify correlations within the system~\cite{uncertainty_product_alon}. Mean-field calculation focuses solely on density distribution, whereas the many-body method accounts for the inter-particle correlations. Thus, the contribution to the variance coming from excited modes is usually a non-zero quantity in many-body calculations but is exactly zero in mean-field methods. Hence, variance is a crucial tool for analyzing the interplay between density distribution and many-body correlations.

\section{Dynamics following rotation quenches}\label{result}
This section explores how a group of repulsively interacting bosons trapped in a two-dimensional anharmonic potential evolves with time after a sudden change in the rotation frequency. Initially, the bosons are  prepared with an initial rotation frequency ($\omega_i$). An abrupt change in the Hamiltonian is then made by lowering the rotation frequency ($\omega_f < \omega_i$) to investigate the dynamical evolution of the system. We divide our findings into three sections. In the first part, the effects of the different magnitude of rotation quenches in a rotationally symmetric trap are discussed. The second part explores how the system evolves in time when the rotational symmetry of the trap is broken by stretching the symmetric trap along the x-direction, referred to as the elongated trap. Finally, the trap is reconstructed in the third part to restore some symmetry of the elongated trap, which allows us to study the rotation quench dynamics in the four-fold symmetric trap potential. The elongated trap exhibits two-fold rotational symmetry, while the four-fold symmetric trap has four-fold rotational symmetry. The dynamics are studied by investigating various quantities such as the time evolution of density, the measure of buildup or loss of coherence in the system, the average angular momentum, and the variances of certain observables. Throughout the calculations, the interaction parameter is set as $\Lambda = \lambda_0 (N-1) = 0.1$. In the mean-field analysis, the dynamical behavior is governed only by the mean-field interaction parameter $\Lambda$. As a result, the particle number $N$ can be as large as relevant while keeping $\Lambda$ constant. In the many-body calculations, we use $N=8$ bosons. This, we shall see below, leads to fragmentation. For the numerical calculations, we utilize a grid consisting of $128 \times 128$ grid points within a box ranging from  $[-8,8) \times [-8,8)$. While a $64 \times 64$ grid is adequate for capturing the dynamics accurately, we choose the higher-resolution to improve the visualization of the density distribution.

\subsection{Rotation quench in a symmetric trap}
\begin{figure*}
    \centering
    \includegraphics[width=0.95\textwidth, angle =-0 ]{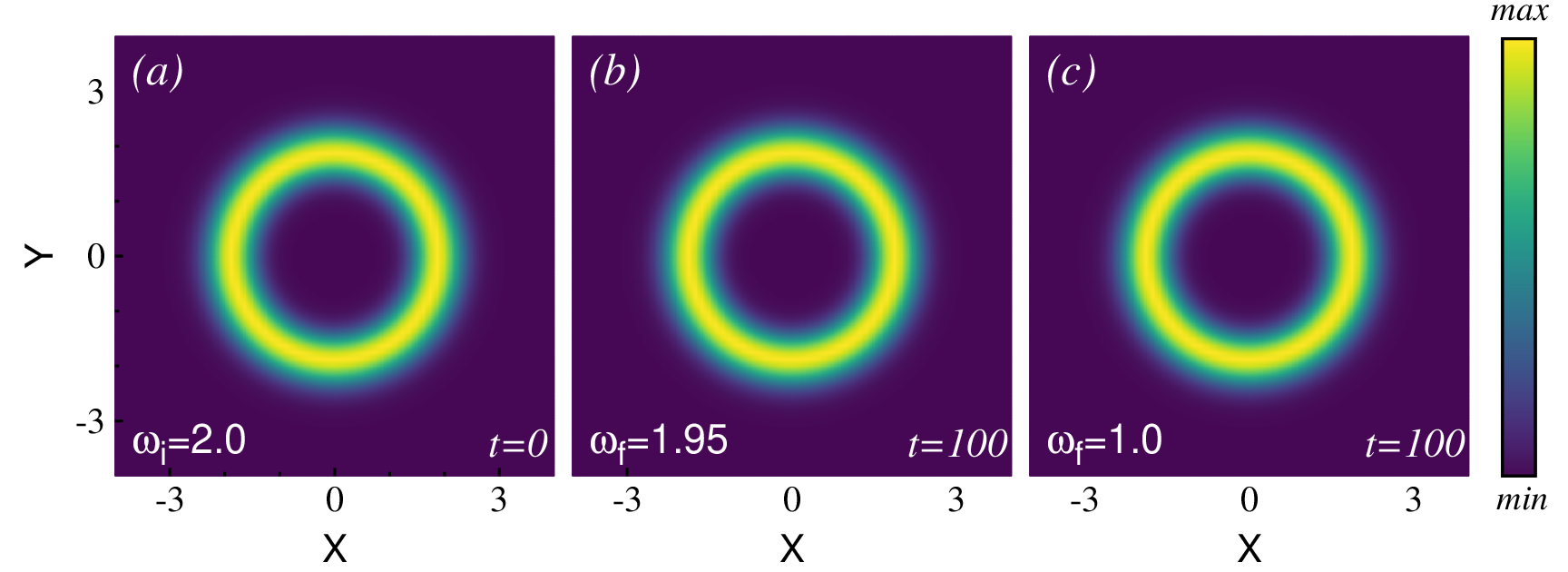}
    \caption{\textbf{Snapshots of one-body density after rotation quenches in the symmetric trap:} (a) The initial density, which takes the shape of a ring due to strong rotation for $\omega_i = 2.0$. (b) The one-body density at time $t=100$ is depicted after the small rotation quench. (c) The density at time $t=100$ is presented after the large rotation quench. All computations are performed for $N=8$ bosons using $M=4$ self-consistent orbitals in the many-body calculation. The interaction parameter is $\Lambda =0.1$. The density pattern remains unchanged during the dynamics, and the results from both the mean-field and many-body calculations show similar behavior. See text for further details. All quantities are dimensionless.}
    \label{fig:density_symmetric}
\end{figure*}

To study the dynamics following a rotation quench in an anharmonic symmetric trap, we consider the following form of the symmetric anharmonic trap, 
\begin{equation}\label{eq.symmetrictrap}
    V(\mathbf{r}) = \frac{1}{4} \left( x^2 + y^2 \right)^2 .
\end{equation}
The initial state is prepared with a rotation frequency of $\omega_i =2.0$. Because of the strong rotation, the initial density distribution takes the form of a ring [see Fig.~\ref{fig:density_symmetric}(a)]. During the quench, the rotation frequency is abruptly changed to a lower value. We explore the dynamics for two different quenches: (i) a small rotation quench ($\omega_f = 1.95$) and (ii) a large rotation quench ($\omega_f = 1.0$). The characterization of the small and large rotation quenches is discussed in the supplementary material. The initial state has the many-body average angular momentum of $\frac{1}{N} \langle \hat{L}_Z \rangle = 7.05$,  which closely aligns with the mean-field calculation.

Fig.~\ref{fig:density_symmetric} shows the post-quench density dynamics obtained from the many-body calculations. After the small rotation quench, the density structure remains unchanged throughout the dynamics up to time $t=200$. A density plot at an intermediate time at $t=100$ is shown in Fig.~\ref{fig:density_symmetric}(b), highlighting its similarity to the initial density. Similar behavior is also observed for the large rotation quench. The density structure remains similar to the initial density [Fig.~\ref{fig:density_symmetric}(c) demonstrates the density distribution at $t=100$ as an illustration]. Both the many-body and mean-field calculations show no difference in the density distribution throughout the time dynamics. The full density dynamics video is provided in the supplementary material. 
  
In terms of the occupation in natural orbitals, the initial state is highly condensed with the first natural orbital population of $\frac{n_1}{N}=0.9987$. After the quench, the occupation remains unchanged throughout the dynamics for both quenches and are identical. Many-body calculations are performed using $M = 4$ orbitals. A more detailed discussion and the convergence of the numerical results are presented in the supplementary material.

Also, the average angular momentum remains constant during the dynamics for both quenches. The initial angular momentum persists throughout the dynamics in both the mean-field and many-body calculations as well. This stagnant behavior of the average angular momentum strongly supports the numerics, since angular momentum is a good quantum number in the symmetric trap under quench. 
We have also calculated the variances of several observables. The qualitative behavior of all variances remains similar for small and large rotation quenches. But there is a mismatch between the mean-field and many-body calculation. A detailed analysis is presented in the Supplemental Material.

So far, to summarize, our study has focused on the rotation quench dynamics in the symmetric anharmonic trap. Throughout the dynamics, the density maintains its initial structure. The state remains highly condensed from the many-body perspective, and no changes are observed in the expectation values of the angular momentum operator during the time evolution. The variance calculations also exhibit uniform behavior. These time-independent behaviors arise from angular momentum being a good quantum number in the symmetric trap. Also, in the rotationally symmetric potential, a rotation quench does not alter the state, which remains an eigenfunction of the quenched Hamiltonian; only the energy of the system is altered. The monotonous behavior of the measured quantities in the symmetric trap motivates us to extend our study of rotation quench dynamics to an asymmetric trap.

\subsection{Rotation quenches in an elongated trap}
In a perfectly symmetric trap, the density evolution, occupation in the natural orbitals, and the average angular momentum show no changes during both strong and weak rotation quenches. Consequently, a pressing inquiry arises: what occurs after a rotation quench when the symmetry of the trap is disrupted, given that angular momentum is no longer a good quantum number in an asymmetric potential. This section investigates the effects of the rotation quenches in an asymmetric trap that is elongated along the x-direction. The form of the trap is chosen as
\begin{equation}\label{eq.elongatedtrap}
    V(\mathbf{r}) = \frac{1}{4} \left( 0.8x^2 + y^2 \right)^2 .
\end{equation}
The initial state is prepared with $\omega_i = 2.0$ and quenched by abruptly reducing the rotation frequency to lower values.  The dynamical study that we performed is threefold: (i) a quench to $\omega_f = 1.95$, referred to as a small rotation quench, where a slight perturbation is introduced to the system; (ii) a quench to $\omega_f = 1.85$, termed an intermediate rotation quench, which injects a moderate yet significant amount of energy into the system; and (iii) a quench to $\omega_f = 1.0$, defined as a large rotation quench. In the supplementary material, we have detailed the characterization process for the three-quench regimes.  The numerical computation is performed with $M=8$ orbitals. The numerical convergence of our results is also presented in the supplementary material. 

The ground state properties in the elongated trap for different rotation frequencies have been explored in Ref.~\cite{sunayana_scirep}. Before proceeding to the dynamics, we first examine the initial state. Because of strong rotation, initially the density breaks into two sub distinct clouds along the x-direction. This trap confinement also breaks the momentum density along the y-direction~\cite{sunayana2024}. The energy per particle for this state is $\frac{E}{N} = -3.4963$, while the initial angular momentum associated with the system is $ \frac{1}{N} \langle \hat{L}_Z \rangle = 11.0891$. The initial state is, of course, fully condensed from the mean-field perspective, and it is completely two-fold fragmented, with $\sim 50\%$ occupation in the first two natural orbitals according to the many-body calculation. Despite these very different descriptions, both systems exhibit the same energy per particle and the same average angular momentum for the initial state. We are curious to see how a system with an identical initial density structure and average angular momentum, observed in both mean-field condensed and many-body fragmented states, evolves over time. We are also interested in exploring the many-body effects that may emerge during the dynamics.

\subsubsection{Dynamical behavior of the density}
\begin{figure*}
    \centering
    \includegraphics[width=0.98\textwidth, angle =-0 ]{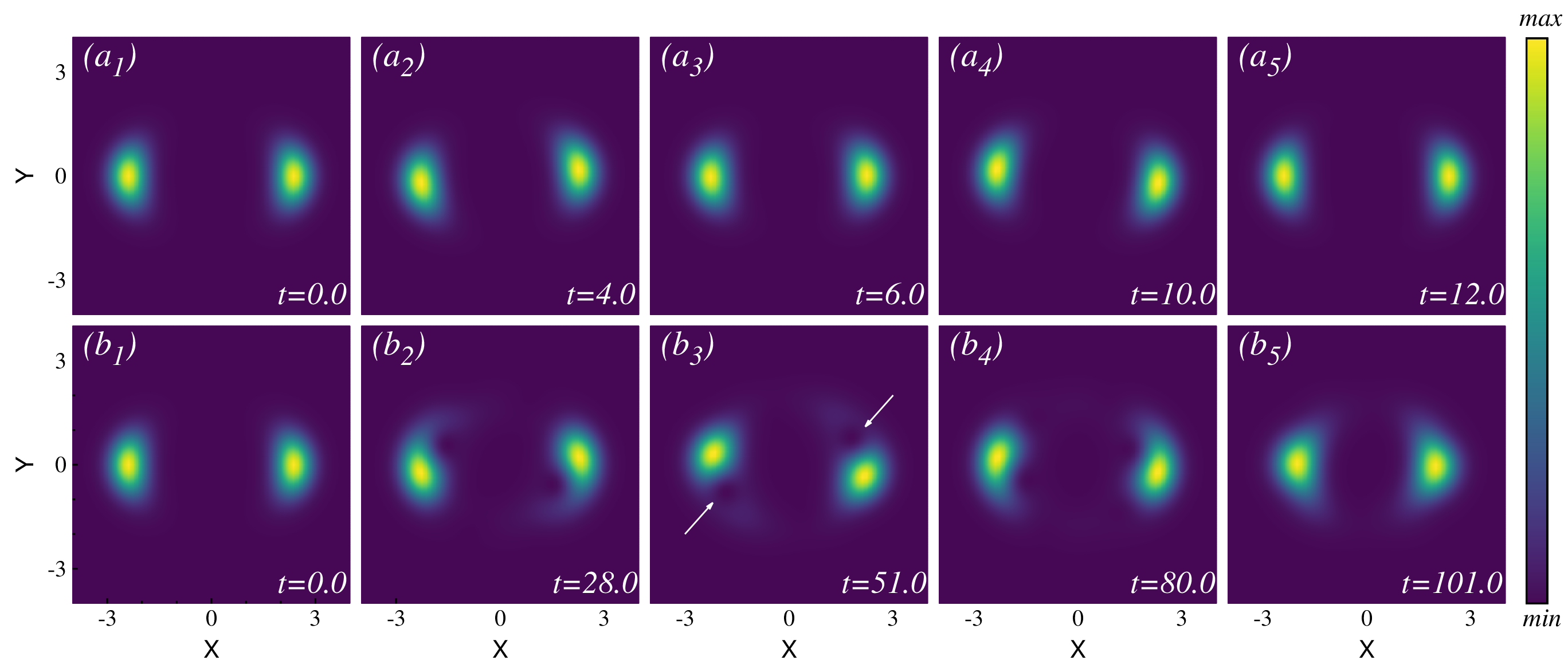}
    \caption{\textbf{Snapshots of the one-body densities after the small and intermediate rotation quenches in the elongated trap.}  ($a_1$) - ($a_5$) illustrate the one-body density over one cycle after the small rotation quench, showing the two distinct density clouds moving along a curved path resembling an arc of a circle. ($b_1$) - ($b_5$) show the one-body density over one cycle after the intermediate rotation quench, with the two separate density clouds exhibiting rotation and oscillation. Throughout the dynamics, one visible vortex appears and disappears in each cloud in a periodic manner. The many-body calculations use $M=8$ self-consistent orbitals. Both the many-body and mean-field calculations (not shown) exhibit similar behavior. For further details, see the main text. The complete video of the one-body density dynamics can be found in the supplementary material. All quantities are dimensionless. }
    \label{fig:density_elongated}
\end{figure*}

First, we concentrate on observing the time evolution of the initially split one-body density in response to the three different strengths of rotation quenches. Fig.~\ref{fig:density_elongated}(upper row) presents the density evolution after the small rotational quench. In this scenario, both the mean-field and many-body calculations exhibit identical dynamical behavior. Therefore, we show snapshots of the density distribution only for the many-body calculations. The initial density distribution is presented in Fig.~\ref{fig:density_elongated}($a_1$). After the small rotation quench, the right cloud starts moving upwards (positive y-direction), while the left cloud moves downwards (negative y-direction). They reach their peak displacements [see Fig.~\ref{fig:density_elongated}($a_2$)] and then move in the reverse direction, returning to their initial positions at time t=6.0 [Fig.~\ref{fig:density_elongated}($a_3$)]. This back-and-forth motion continues, with the clouds reaching their maximum displacements in the opposite direction at time t=10.0 [Fig.~\ref{fig:density_elongated}($a_4$)]. At time $t=12$, the two density clouds again pass through the initial position [Fig.~\ref{fig:density_elongated}($a_5$)]. This oscillation pattern persists over time with a period of $t \sim 12$. The density clouds do not travel in a straight line; instead, they oscillate back and forth along a curved trajectory. The full density dynamics movie is available in the supplementary material.

In the intermediate rotation quench, a moderate amount of energy and angular momentum are injected into the system (see supplementary material for details), but the quench is not sufficient for the two initially separated clouds to interact with each other and complete a full rotation. The mean-field and many-body calculations yield nearly identical density dynamics in this case as well. Therefore, the snapshots of the many-body density distribution at various times are presented in Fig.~\ref{fig:density_elongated}(bottom row). Initially, the two distinct density clouds separated along the x-direction oscillate vertically. The early time dynamics (up to $t=11$) resembles the density dynamics observed in the small rotation quench, but with a larger amplitude of oscillations. In contrast to the small rotation quench where the oscillations persist throughout the time, a more complex dynamics is observed in this scenario. As time progresses beyond $t>25$, a density minimum (vortex) emerges alongside each density cloud and becomes more pronounced at $t=51$ [Fig.~\ref{fig:density_elongated}($b_3$)]. An intriguing dynamics unfolds as the two sub clouds start rotating around their own axis, centering the two minima of the potential. This introduces a second type of motion to the system, where the clouds oscillate along an arc in the y-direction while simultaneously undergoing internal rotation. The vortices themselves also rotate in a synchronized way with the clouds. We have analyzed the phase profile of the wave function to pinpoint the vortex position, as detailed in the supplementary material. A full density and vortex dynamics video is available in the supplementary video. The two density clouds never intersect with each other during their motion. The system exhibits the second kind of motion until about $t \sim 99$. Afterward, the system transits back to the first kind of motion, i.e., oscillating along the arc in the y direction [Fig.~\ref{fig:density_elongated}($b_5$)]. The system alternates between these two types of motions in a repeating cycle throughout the dynamics. The emergence of the visible vortices appear to balance the angular momentum. Later, we discuss the angular momentum dynamics and establish a relation between the vortex dynamics and the corresponding angular momentum.

\begin{figure*}
    \centering
    \includegraphics[width=0.98\textwidth, angle =-0 ]{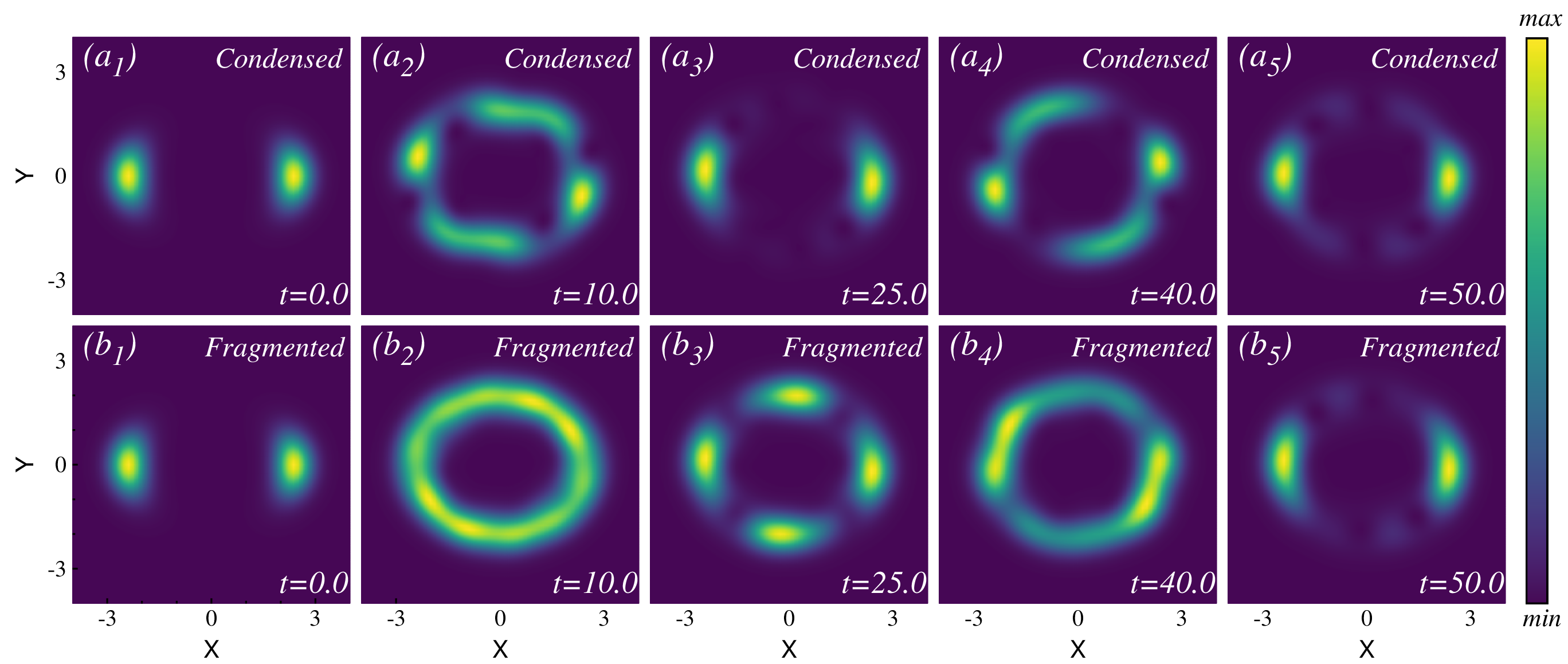}
    \caption{\textbf{Snapshots of the one-body density evolution after the large rotation quench in the elongated trap.}  ($a_1$) - ($a_5$) depict the one-body density at different times for the mean-field calculation, which is a condensed state by definition. ($b_1$) - ($b_5$) show the one-body density at the same corresponding times for the many-body calculation, which is a fragmented state. In this large rotation quench, the initial two separate clouds now interact with each other and are capable of completing a full circular rotation. The mean-field and the many-body densities exhibit notable differences, not only in terms of the timescale but also in the structure of the density itself. See the text for further discussions. The complete video of the density dynamics is available in the supplementary material. All quantities are dimensionless. }
    \label{fig:density_elongated_1.0}
\end{figure*}

The density evolution after the large rotation quench is presented in Fig.~\ref{fig:density_elongated_1.0}. The mean-field results are shown in the upper row, and the many-body results are displayed in the lower row. Given the strength of the rotation quench, many excited states actively contribute to the dynamics. Consequently, in the dynamics, significant differences arise between the mean-field condensed state and the many-body fragmented state. Due to the strong rotation kick, the two initially well-separated clouds are now capable of rotating in a circular path. The resulting circular rotating cloud exhibits distinct structures as it evolves over time. 
For the condensed system, the density distribution exhibits a circular-like density pattern at $t=10.0$ [Fig.~\ref{fig:density_elongated_1.0}($a_2$)]. In the intermediate time, four humps are observed in the density distribution (not shown). One can see the full density dynamics movie available in the supplementary material. Eventually, the clouds return to a state resembling its initial configuration [Fig.~\ref{fig:density_elongated_1.0}($a_3$)]. The density evolution starts with two distinct clouds separated along the x-direction and gradually returns to its initial shape during the dynamics, defining one complete cycle. In this particular case, the time taken for one complete cycle is $t\sim 25$. Again, at time $t\sim 50$, the cloud returns to a state resembling the initial configuration after passing through all the intermediate states [Fig.~\ref{fig:density_elongated_1.0}($a_5$)]. 
For the fragmented system, the two initially well-separated clouds commence rotation, forming a circular ring [Fig.~\ref{fig:density_elongated_1.0}($b_2$)]. During the evolution, while the condensed state shows four distinct humps in the density, the many-body fragmented state is still in the process of achieving this structure (not shown in the figure). Later, when the mean-field condensed state density returns to a configuration similar to the initial state, the fragmented state  achieves the four-hump density structure [Fig.~\ref{fig:density_elongated_1.0}($b_3$)]. As time progresses, the density cloud eventually returns to a state closely resembling its initial configuration [Fig.~\ref{fig:density_elongated_1.0}($b_5$)]. From the initial two-hump density to the four-hump density and back to an initial-like density structure, the many-body fragmented system dynamics takes approximately $t \sim 50$, precisely twice the time required by the mean-field condensed system. During the dynamics, the many-body and the mean-field calculations pass through similar density structures, but the many-body fragmented system shows more uniform density distributions in each case.
There are four distinct density structures observed during the oscillation: (i) initially two well separated density clouds; (ii) a circular like density structure; (iii) four distinct humps in the density; and (iv) the density cloud eventually returns to a state closely resembling its initial density. 
We have performed additional many-body calculations for $N = 10$ and $N = 12$ bosons, and the results closely resemble those obtained for $N = 8$. We have interlinked each structure of the density dynamics with the corresponding angular momentum dynamics in the angular momentum section.

\begin{figure*}
    \centering
    \includegraphics[width=1.0\textwidth, angle =-0 ]{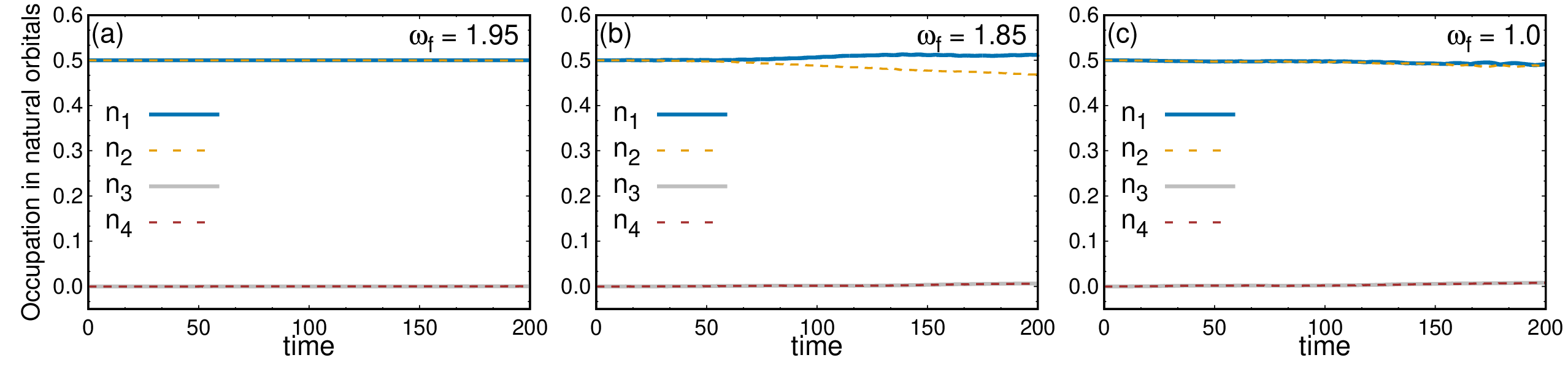}
    \caption{\textbf{Natural orbital occupations for different rotation quenches in the elongated trap.} The occupations in the first four natural orbitals are presented for (a) small, (b) intermediate, and (c) large rotation quenches. All calculations are conducted using $M=8$ orbitals. The initial state is two-fold fragmented, with $\frac{n_1}{N} \approx 0.5$ and $\frac{n_2}{N} \approx 0.5$, while the remaining two orbitals contribute insignificantly. For small rotation quench, the initial two-fold fragmented state persists throughout the time. After the intermediate rotation quench, the initial two-fold fragmented state shows a tendency towards the buildup of coherence. After the large rotation quench, the initial two-fold fragmented state remains almost two-fold fragmented throughout the time dynamics, similar to the small rotation quench. All quantities shown are dimensionless.}
    \label{fig:occupation_elongated}
\end{figure*}

\subsubsection{Dynamics of the natural orbital occupations}
The initial state in the elongated trap is a two-fold fragmented state, with occupation in the first four natural orbitals being $\frac{n_1}{N} \simeq \frac{n_2}{N} \simeq 0.5$ and $\frac{n_3}{N} \sim \frac{n_4}{N} \sim 10^{-5}$. Fig.~\ref{fig:occupation_elongated} illustrates the time evolution of occupations in the first four natural orbitals for the three different quenches. In the small rotation quench, the initial two-fold fragmented state remains stable throughout the dynamics.  The first two natural orbitals, $\frac{n_1}{N}$ and $\frac{n_2}{N}$, having occupations of $\sim 50\%$, overlap with each other, while the occupations of the other two natural orbitals are negligible [see Fig.~\ref{fig:occupation_elongated}(a)].

In the intermediate rotation quench, the first two natural orbitals of the initial two-fold fragmented state  undergo changes as presented in Fig.~\ref{fig:occupation_elongated}(b). Over time, $\frac{n_1}{N}$ increases and $\frac{n_2}{N}$ decreases from initial $50\%$ population. This observation illustrates a distinct type of dynamics, indicating the tendency to develop coherence in the system. Note that the occupation numbers are computed by diagonalizing the reduced one-body density matrix (Eq.~\ref{rdm}), which encapsulates information about the full many-body wave function. Consequently, the occupation numbers are global properties of the system and do not correspond to individual clouds separately.

For the large rotation quench [see Fig.~\ref{fig:occupation_elongated}(c)], the behavior of occupation in the natural orbitals appears similar to that of the small quench. The first two natural orbitals, $\frac{n_1}{N}$ and $\frac{n_2}{N}$, overlap with each other during the dynamics. However, at larger times, slight fluctuations are observed although the state remains almost two-fold fragmented. Thus, from the many-body perspective, both the small and large quenches exhibit similar behavior in terms of occupation in the natural orbitals, but their density dynamics are markedly different. Consequently, further studies are required to explore the dynamics of these three types of quenches, see below. 

\subsubsection{Time evolution of the average angular momentum }
\begin{figure*}
    \centering
    \includegraphics[width=1.0\textwidth, angle =-0 ]{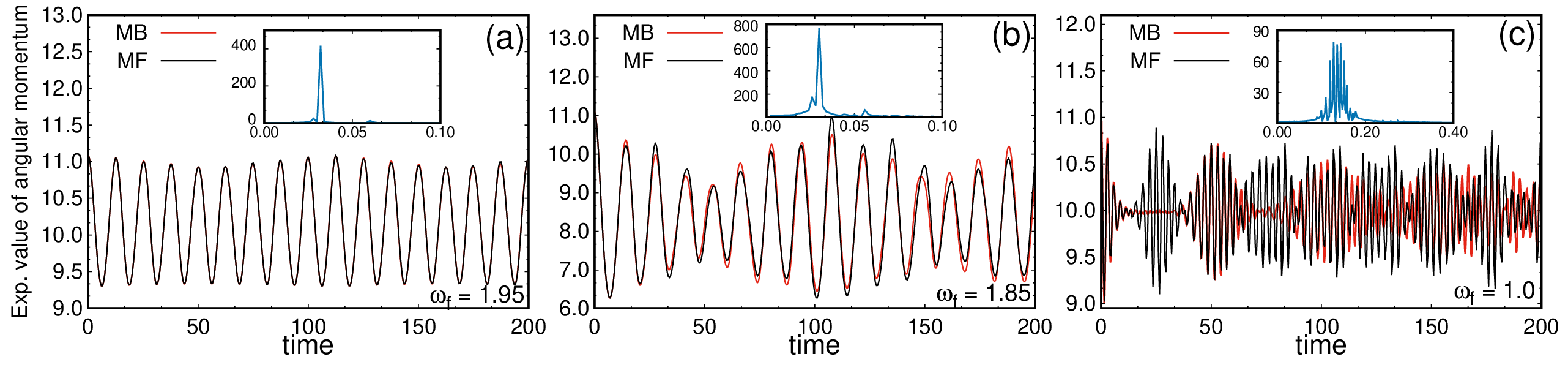}
    \caption{\textbf{Angular momentum expectation value per particle [$\frac{1}{N}\langle \Psi \vert \hat{L}_Z \vert \Psi \rangle$] after rotation quenches in the elongated trap.}  The analysis is performed for (a) small, (b) intermediate, and (c) large rotation quenches. The mean-field results employing one time-adaptive orbital ($M = 1$) are illustrated in black, and the corresponding many-body results utilizing $M = 8$ time-adaptive orbitals are shown in red. The insets provide the Fourier transform of the many-body angular momentum over time, revealing the number of rotational modes involved in the dynamics. All quantities shown are dimensionless.}
    \label{fig:angular_elongated}
\end{figure*}

Angular momentum is a fundamental property in two-dimensional systems, especially significant when rotation is present in the system. It is important to note that the angular momentum is not a conserved quantity in the elongated trap, unlike in the symmetric trap. In Fig.~\ref{fig:angular_elongated}, we plot the expectation value of the angular momentum operator for three different quenches. All results are computed for mean-field condensed systems as well as many-body fragmented systems. The initial state has the average angular momentum of $\frac{1}{N}\langle \hat{L}_Z \rangle = 10.95$, which is consistent in both the mean-field and many-body calculations. After the small rotation quench [Fig.~\ref{fig:angular_elongated}(a)], the angular momentum dynamics show periodic oscillation with a lower bound of $9.3$. The angular momentum dynamics of the mean-field condensed state and the many-body fragmented state overlap throughout the time. We made the Fourier transform of the angular momentum vs. time data and found that the oscillation has only a single dominant frequency [see inset of Fig.~\ref{fig:angular_elongated}(a)]. The Fourier transformation of angular momentum dynamics gives information about the number of rotational modes present during the dynamics. Thus, for the small rotation quench, only a single rotational mode is present. When comparing the single rotational mode to the density evolution, this mode manifests as the oscillation of the density clouds along an arc in the y-direction.

In the intermediate rotation quench, an oscillatory behavior is also observed, but the pattern differs from that observed in the small rotation quench. The oscillation pattern in Fig.~\ref{fig:angular_elongated}(b) shows a beating oscillation, with gradual attenuation, resurgence, and subsequent damping. We explore how the damping of angular momentum relates to the dynamics of the density. In the density evolution, two distinct types of dynamics are observed. First, there is a persistent oscillation in an arc path in the y direction, that remains consistent over time. 
Second, another type of motion emerges shortly after the quench, characterized by the formation of visible vortices and the rotation of those vortices and sub clouds around the potential minima. This second type of motion appears at $t > 25$ and contributes to minimizing the average angular momentum. 
To understand why the angular momentum oscillates with a changing amplitude, we studied how the vortices move after the quench. We found that in our system, the oscillation happens because the vortices change their position relative to the center of rotation. On the top, the change in the amplitude of the oscillations is linked to appearance of visible vortices inside the density clouds and the deformation of the cloud. A more detailed explanation can be found in the supplementary material. 
However, at $t > 100$, both the vortex motion and the associated rotational motion disappear, leading to a subsequent increase in the average angular momentum once again. The mean-field and the many-body calculations are qualitatively the same in nature but slightly deviate quantitatively. The Fourier analysis of the many-body angular momentum gives two major frequencies [inset of Fig.~\ref{fig:angular_elongated}(b)]. This also reinforces the existence of two types of dynamics associated with this rotation quench. 

For the large rotation quench, initially, the results both from the mean-field and many-body calculations deviate significantly. Fig.~\ref{fig:angular_elongated}(c) shows the average angular momentum after the large rotation quench, exhibiting beating oscillations and forming a periodically-appearing wave-packet shape. Notably, the mean-field dynamics pattern repeats twice as often as the many-body dynamics. There is a one-to-one correspondence between the density evolution (see Fig.~\ref{fig:density_elongated_1.0}) and the average angular momentum dynamics, as the time period of the many-body density dynamics is twice that of the mean-field density dynamics. Also the circular shape in the density pattern corresponds to the lower values of the angular momentum, while two distinct density clouds result in the higher values of the angular momentum. The Fourier analysis [inset of Fig.~\ref{fig:angular_elongated}(c)] shows multiple peak, which indicates the presence of several rotational modes in the dynamics. \\

Our analysis of both small and large rotation quenches reveals that, from the many-body perspective, the system remains fully two-fold fragmented throughout the dynamics, whereas from the mean-field perspective, of course, it remains condensed. However, while the density and angular momentum dynamics of the mean-field condensed state and the many-body fragmented state show similar behavior for the small rotation quench, they exhibit clear differences for the large rotation quench. In mean-field calculations, the full state is modeled as the sum of the two states (left and right density clouds), while in the many-body calculations, the state is represented as the product of the two. When there is practically no interaction between the left and right clouds---meaning that both the clouds evolve independently---the mean-field and the many-body methods yield similar dynamics of the density and angular momentum, which is observed in the small rotation quench. However, in the large rotation quench, there is significant interaction between the left and right clouds during the dynamics. As a result, the mean-field condensed state and the many-body fragmented state are expected to produce different dynamics. This is indeed the case. 

\begin{figure*}
    \centering
    \includegraphics[width=0.65\textwidth, angle =-0 ]{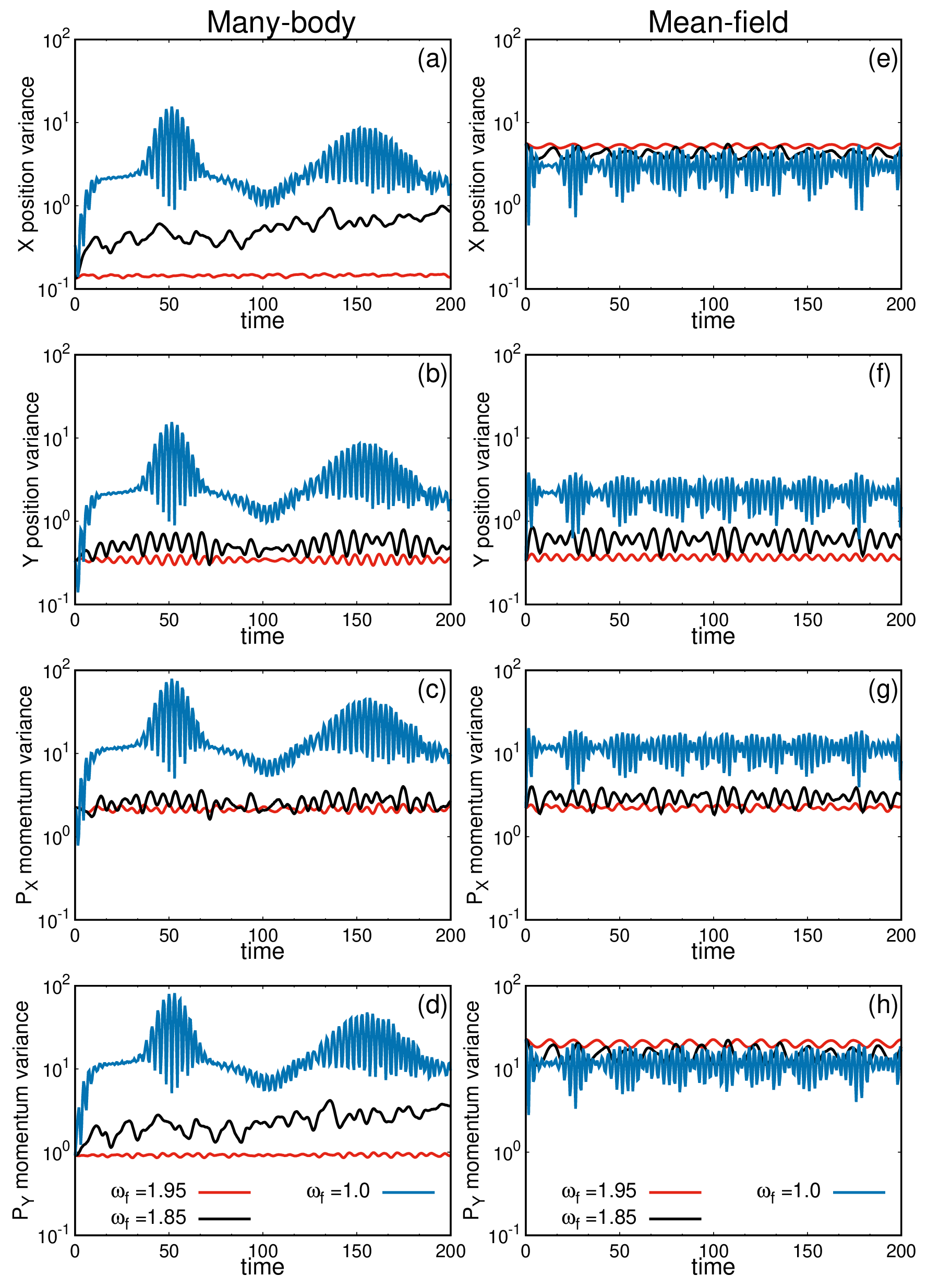}
    \caption{\textbf{Dynamics of the position, and momentum variances for different rotation quenches in the elongated trap.} Time evolution of position variances, $\frac{1}{N} \Delta_{\hat{X}}^2 (t)$ (a,e) and $\frac{1}{N} \Delta_{\hat{Y}}^2 (t)$ (b,f), and momentum variances, $\frac{1}{N} \Delta_{\hat{P}X}^2 (t)$ (c,g) and $\frac{1}{N} \Delta{\hat{P}_Y}^2 (t)$ (d,h), for the many-body (left column) and the mean-field (right column) calculations. The many-body calculations are carried out using $M=8$ time adaptive orbitals. There are substantial differences between the mean-field and many-body results, which become more prominent with increase in the quench magnitude. See text for further details. All quantities shown  are dimensionless.}   
    \label{fig:variance_elon}
\end{figure*}

\subsubsection{Many-body variances}
The variances provide an effective method for detecting even a small amount of correlations within a system. This tool naturally serves to quantify the deviation of the system from being in an eigenstate of the observable quantities~\cite{variance_ofir2015,uncertainty_product_alon}.
To understand the impact of a sudden change in rotation on the system, we analyze the time evolution of the variance of different observables, offering insights into post-quench behavior in the elongated trap.

In Fig.~\ref{fig:variance_elon}, we plot the variances of different quantities. The left column depicts the many-body variances, and the right column presents the corresponding mean-field variances for all three different rotation quenches. It is crucial to note that in the mean-field calculation, we are computing the variance of observables for a condensed system, where only a single ($M=1$) orbital is considered. In contrast, the many-body calculation treats the system as fragmented. Thus, in this discussion, we have compared their respective dynamical behaviors side by side. First, the time evolution of the position variance per particle along the x-direction is explored in Fig.~\ref{fig:variance_elon}(a, e). In the many-body calculation, the initial value of $\frac{1}{N} \Delta_{\hat{X}}^2 (t)$ is $0.13$, whereas it is $5.57$ in the mean-field calculation. This large difference arises because we utilize an elongated trap, causing the density cloud to split into two in the x-direction. The many-body variance exhibits a lower value due to the fully fragmented state, whereas the mean-field calculation describes the state as condensed, resulting in a higher value for $\frac{1}{N} \Delta_{\hat{X}}^2 (t)$~\cite{sunayana_scirep}. Throughout the time dynamics, both mean-field and many-body variances exhibit oscillations with small amplitudes. For the intermediate rotation quench, the pattern is slightly different. The variances oscillate in an aperiodic manner. As time progresses, the mean-field oscillates within the same upper and lower bounds, while the many-body $\frac{1}{N} \Delta_{\hat{X}}^2 (t)$ starts to increase. This increase is attributed to the buildup of coherence in the many-body calculation, which is reflected in the occupation of the natural orbitals [see Fig.~\ref{fig:occupation_elongated}(b)]. In the large rotation quench, the difference between mean-field and many-body results becomes very prominent. The many-body $\frac{1}{N} \Delta_{\hat{X}}^2 (t)$ exhibits a higher bound of $\sim 30$, while the mean-field $\frac{1}{N} \Delta_{\hat{X}}^2 (t)$ shows oscillations with an upper bound of $\sim 6.0$. 

Continuing, we examine the time evolution of the position variance per particle along the y-direction [Fig.~\ref{fig:variance_elon}(b), (f)]. The initial value of $\frac{1}{N} \Delta_{\hat{Y}}^2 (t)$ in both many-body and mean-field calculations is approximately $0.33$. For the small rotation quench, the mean-field result exhibits regular oscillations, while the many-body result oscillates with changing magnitudes. This oscillation pattern becomes aperiodic as the strength of the rotation quenches increases. The difference between the mean-field and many-body results is already significant in the intermediate rotation quench and becomes higher in the case of the large rotation quench. 

Next, we examine the behavior of the momentum variance per particle $\frac{1}{N} \Delta_{\hat{P_X}}^2 (t)$ along the x-direction [Fig.~\ref{fig:variance_elon}(c), (g)]. The behavior of $\frac{1}{N} \Delta_{\hat{P_X}}^2 (t)$ mirrors that of $\frac{1}{N} \Delta_{\hat{Y}}^2 (t)$. Here as well, the oscillation pattern becomes aperiodic with an increase in the strength of the rotation quenches. However, both the upper and lower bounds, as well as the difference between the mean-field and many-body results, increase substantially from the small rotation quench to the large rotation quench. 

Now let us delve into another interesting result: the time dynamics of the momentum variance per particle $\frac{1}{N} \Delta_{\hat{P}_Y}^2 (t)$ along the y-direction [Fig.~\ref{fig:variance_elon}(d), (h)]. The initial value of $\frac{1}{N} \Delta_{\hat{P}_Y}^2 (t)$ in the many-body calculation is $0.91$, while it is $22.68$ in the mean-field calculation. There is a significant difference between mean-field and many-body $\frac{1}{N} \Delta_{\hat{P}_Y}^2 (t)$, primarily because the elongated trap in the x-direction in the presence of strong rotation splits the spatial density in the x-direction and the momentum density in the y-direction~\cite{sunayana2024}. Thus, we observe a similar kind of anisotropy in $\frac{1}{N} \Delta_{\hat{P}_Y}^2 (t)$ to that in $\frac{1}{N} \Delta_{\hat{X}}^2 (t)$. For the small rotation quench, both the mean-field and many-body $\frac{1}{N} \Delta_{\hat{P}_Y}^2 (t)$ oscillate periodically. In the case of the intermediate rotation quench, there is no periodic-like oscillation. The many-body $\frac{1}{N} \Delta_{\hat{P}_Y}^2 (t)$ starts to increase with time as the system deviates from its initial 50\% fragmented state. Finally, for the large rotation quench, substantial differences are observed in $\frac{1}{N} \Delta_{\hat{P}_Y}^2 (t)$ between the mean-field and many-body results.

Another notable observation is that the position and momentum variances along the x- and y-directions appear similar for the large rotation quench. This is attributed to the substantial amount of energy and angular momentum being pumped into the system, causing a large number of  excited states to participate in the dynamics. Because of the significant rotational kick, the dominant influence on the system arises from the rotation itself rather than the shape of the trap.\\

The rotation quench in the elongated trap proves to be more intriguing compared to the symmetric trap. The dynamics of almost all measured observables exhibit distinct patterns depending on the magnitude of the rotation quench. The mean-field condensed and the many-body fragmented states exhibit differences across most observables. These differences are minimal for small rotational quenches but become more pronounced as the magnitude of the rotational quench increases. The initial state is prepared under high rotation, which induces an effective double-well structure, resulting in the splitting of the density into two distinct parts. Quenching the rotation to a lower frequency effectively alters the trap geometry. This type of analysis is also applied to other systems~\cite{Ofir_PRA_R}. As the magnitude of the rotation quench increases, the density lobes tend to merge. Notably, there is no sharp transition between different quench regimes, indicating a continuous crossover in the dynamics.

\subsection{Rotation quenches in a four-fold symmetric trap}
Our observations so far demonstrate that in the perfectly symmetric trap, none of the examined quantities show significant change during their time evolution after the rotation quench, regardless of the strength of the quench. In the elongated trap, symmetry breaking leads to distinct dynamics across several observables. Motivated by the complex time dynamics observed in the elongated trap, we extend our investigation using a four-fold symmetric trapping potential. We aim to study the effects of partially restoring symmetry along the y-direction with respect to the elongated trap, which may potentially cause the disappearance of certain modes. Furthermore, we are also interested to explore whether certain many-body effects present (absent) in the dynamical study of the elongated trap might disappear (appear) in the four-fold trap geometry. To study the rotation quench dynamics in a four-fold symmetric potential, the form of the potential is chosen as
\begin{equation}
    V(\mathbf{r}) = \frac{1}{4} \left( x^4 + y^4 \right).
\end{equation}
The initial state is prepared with $\omega_i = 2.2$. The initial rotation frequency is a bit higher compared to the elongated trap, as we focus on starting the dynamics from a fully four-fold fragmented state from the many-body perspective. We study the rotation quench dynamics for three distinct regimes: (i) a small rotation quench ($\omega_f = 2.15$), (ii) an intermediate rotation quench ($\omega_f = 2.0$), and (iii) a large rotation quench ($\omega_f = 1.6$). The characterizations of small, intermediate, and large rotation quenches are thoroughly described in the supplementary material. Because of the strong initial rotation, the density splits into four distinct sub clouds [see Fig.~\ref{fig:density_four_2.0}($a_1$)]. The energy per particle stands at $\frac{E}{N} = -7.9661$, with an average angular momentum of $\frac{1}{N} \langle \hat{L}_Z \rangle = 19.4978$. Although the mean-field and the many-body methods describe different states, they both produce essentially the same ground-state energy per particle, the same density profile, and the same average angular momentum for the initial state.

\subsubsection{The time-dependent density}
We initiate the analysis by examining the evolution of the one-body density. In the small rotation quench, the system is perturbed with a very slight rotation change. As a result, the four distinct clouds begin to rotate independently in the anti-clockwise direction, with the axis of rotation passing through the center of their potential minima. However, the clouds do not interact with each other; instead, they maintain their individual circular motion (not shown in the main text). We have provided the full density dynamics video in the supplementary material. This behavior persists until $t=200$. Both the mean-field and many-body dynamics yield the same behavior. 

\begin{figure*}
    \centering
    \includegraphics[width=1\textwidth, angle =-0 ]{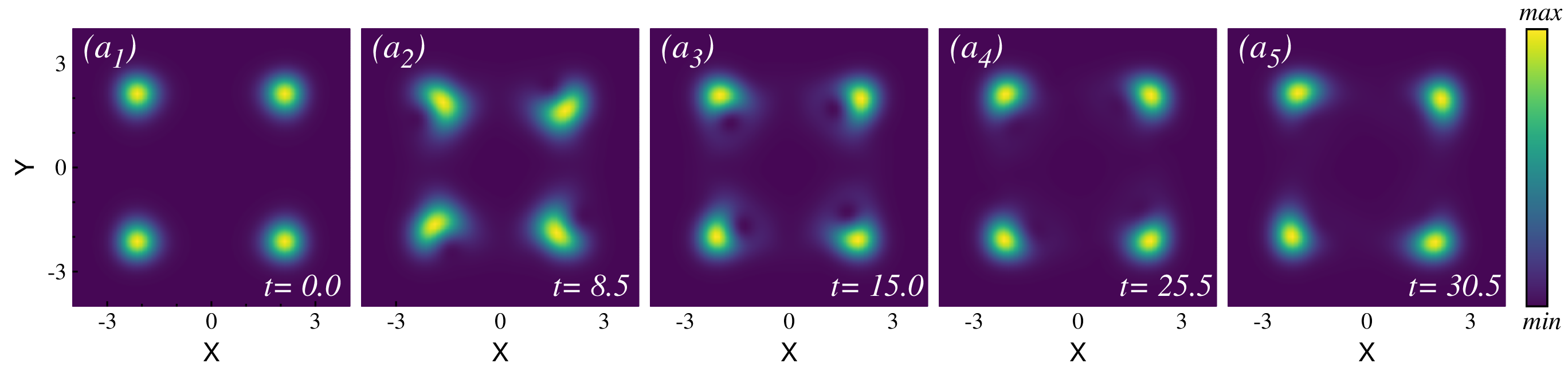}
    \caption{\textbf{Snapshots of the density evolution after the intermediate rotation quench in the four-fold symmetric trap.}  The many-body calculation is obtained using  $M=12$ time-adaptive orbitals.  The initial four well-separated density sub clouds ($a_1$) begin to rotate, centering around the potential minimum. During the dynamics, one visible vortex in each sub cloud appears. These vortices fade in ($a_2$), become prominent in ($a_3$), and then fade away again in ($a_4$) and completely disappear in ($a_5$). Two distinct dynamics are evident here: (i) rotation of each cloud without any visible vortex and (ii) rotation of each cloud, including the visible vortex. The full density dynamics movie is available in the supplementary material. Qualitatively, the results from both the many-body and mean-field calculations show good agreement (mean-field results not shown). See text for further details. All quantities shown are dimensionless.}
    \label{fig:density_four_2.0}
\end{figure*}

For the intermediate rotation quench, the density dynamics become more complex. Snapshots of the density at various time points are illustrated in Fig.~\ref{fig:density_four_2.0}, computed using the many-body description. The mean-field and the many-body density dynamics exhibit the same behavior in this case as well. The initial dynamics resemble those of the small rotation quench---the initially well-separated density clouds begin to rotate along the axis, centering their potential minima up to time $t=4$. As time progresses, one visible vortex appears in each sub cloud [see Fig.~\ref{fig:density_four_2.0}($a_2$)]. We have analyzed the phase profile of the wave function to pinpoint the vortex position, as detailed in the supplementary material. A full vortex dynamics video is available in the supplementary video.
These vortices also rotate along with the density sub clouds. Throughout this process, no two density clouds ever touch each other; instead, they rotate individually. The vortices become more prominent in the course of time at $t=15.0$ [see Fig.~\ref{fig:density_four_2.0}($a_3$)]. As time progresses, the vortices start to disappear [see Fig.~\ref{fig:density_four_2.0}($a_4$)] and completely disappear for $t \sim 30.5$ [see Fig.~\ref{fig:density_four_2.0}($a_5$)]. We have illustrated just one cycle of the density dynamics up to time $t=30.5$. During the full course of time evolution, the vortices periodically emerge and disappear. See the full dynamics video in the supplementary material. The corresponding angular momentum exhibits lower and higher values depending on whether a visible vortex is present in each sub-cloud, which we discuss in detail in the angular momentum section. 

\begin{figure*}
    \centering
    \includegraphics[width=0.95\textwidth, angle =-0 ]{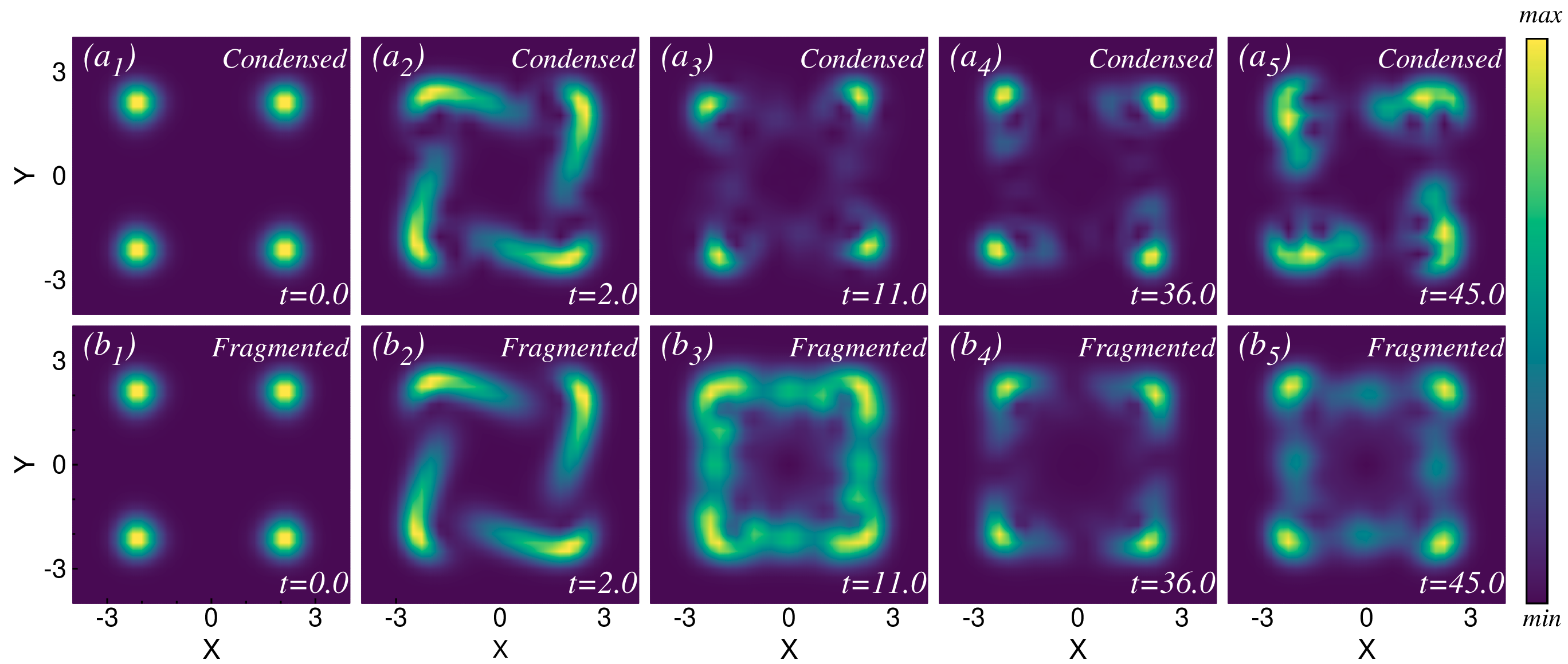}
    \caption{\textbf{Snapshots of the one-body density evolution after the large rotation quench in four-fold symmetric trap.} ($a_1$) - ($a_5$) depict the one-body density at different times for the mean-field calculation. ($b_1$) - ($b_5$) show the one-body density at the same times for the many-body calculation. Many-body results are obtained for $N=4$ bosons and $M = 28$ time-adaptive orbitals. In this quench, the four distinct clouds interact, forming various interference patterns and generating multiple vortices over time. The mean-field and the many-body densities exhibit significant differences in both their timescales and density structures during the dynamics. Further details are provided in the text, and a full video of the density dynamics is available in the supplementary material. All quantities shown are dimensionless.}
    \label{fig:density_1.6}
\end{figure*}

In the large rotation quench, we inject a substantial amount of energy into the system. In this quench, many excited states actively participate in the dynamics, making the system behave in more complicated ways. The substantial rotational kick causes the four distinct density clouds to collide with each other, resulting in the formation of interference patterns. To accurately describe the dynamics up to time $t=200$, a large Hilbert space is required. To present a converged result, we need to reduce the number of particles to $N=4$, which allows us to raise the number of orbitals to $M=28$ in our computation. 
The mean-field density dynamics is shown in the upper panel of Fig.~\ref{fig:density_1.6}, while the many-body results are displayed in the lower panel. Significant differences between the many-body and mean-field results are observed during the dynamics. Because of the strong rotational kick, the initially four well-separated clouds interact with each other and acquire the capacity to rotate in a counterclockwise direction. But, during the time evolution of the density, no such pattern or period is observed. We showcase snapshots at different time points (up to $t=45$) for both many-body and mean-field calculations in order to highlight the difference in the density patterns between the two methods. The complete density dynamics video is provided in the supplementary material. During the dynamics, except for the initial state, the many-body fragmented state exhibits a more dispersed density profile compared to the mean-field condensed state. Also, the mean-field dynamics display prominent multiple vortex configurations, whereas the vortices appear diminished in the many-body density at the same time.

\subsubsection{Dynamics of the natural orbital occupation}
\begin{figure*}
    \centering
    \includegraphics[width=0.88\textwidth, angle =-0 ]{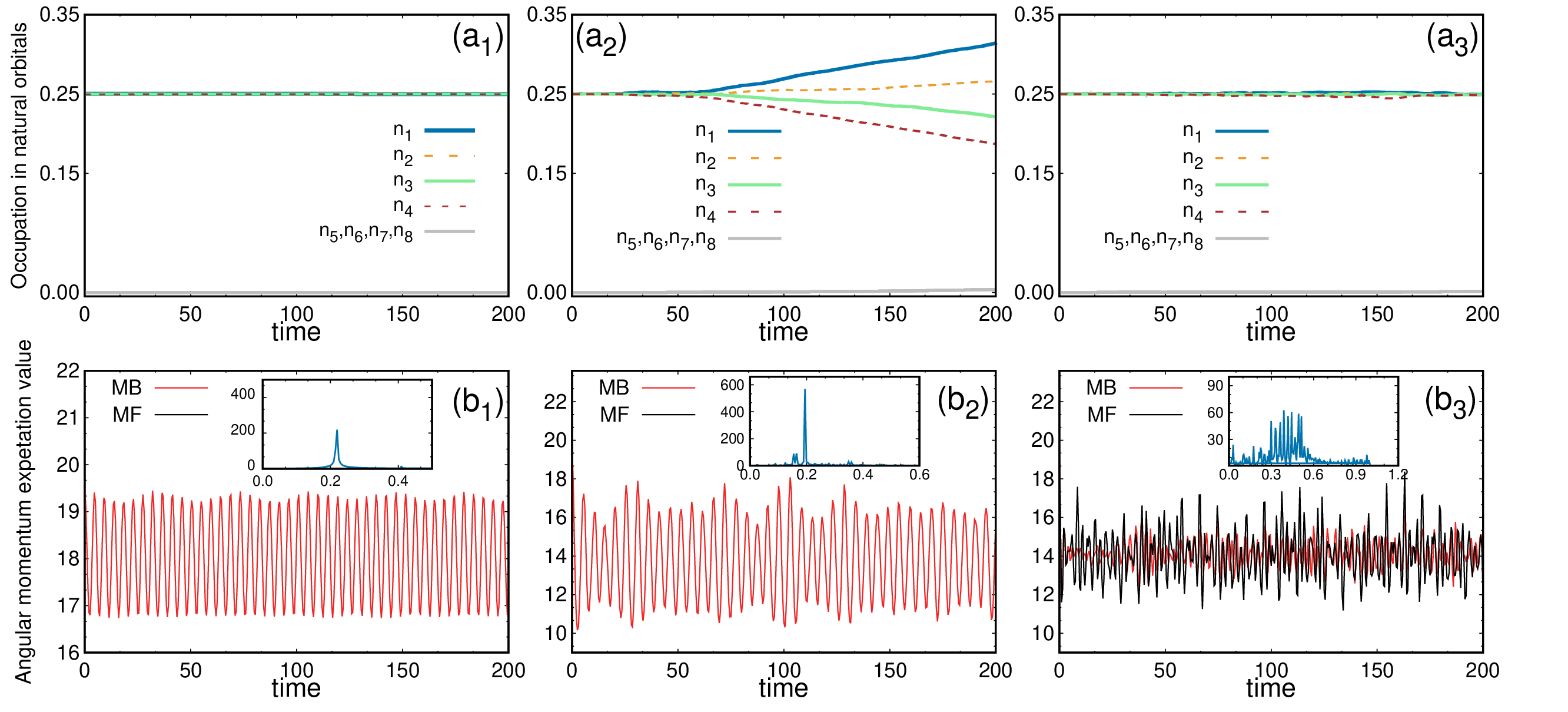}
    \caption{\textbf{Occupations in the natural orbitals and the average angular momentum in the four-fold symmetric trap.} Upper row: occupations in the first eight natural orbitals are presented. The initial state exhibits four-fold fragmentation ($n_{1,2,3,4}/N \approx 0.25$). ($a_1$) Small rotation quench: the four-fold fragmented state persists throughout the dynamics. ($a_2$) Intermediate rotation quench: the initial four-fold fragmented state showing a tendency to buildup of coherence. ($a_3$) Large rotation quench: the four-fold fragmented state persists throughout the dynamics, though a slight fluctuation is observed.
    Lower row: angular momentum expectation value per particle, $\frac{1}{N}\langle \Psi \vert \hat{L}_Z \vert \Psi \rangle$. The mean-field (black) and the many-body (red) calculations are shown for ($b_1$) small, ($b_2$) intermediate and ($b_3$) large rotation quenches.  The insets show the Fourier transform of the many-body angular momentum vs time to extract the number of rotational modes present in the dynamics.  All calculations are conducted using $M=12$ orbitals for small and intermediate rotation quench and $M=28$ orbitals for large rotation quench. All quantities shown are dimensionless. }
    \label{fig:occupation_four}
\end{figure*}

The initial state is a fully four-fold fragmented state with $\sim 25\%$ occupation in the first four natural orbitals. Fig.~\ref{fig:occupation_four}(upper row) illustrates the evolution of the  natural orbitals under small, intermediate and large rotation quenches. 
Fig.~\ref{fig:occupation_four}($a_1$) presents the time evolution of the occupations of the first eight natural orbitals following the small rotation quench. The initial four-fold fragmented state remains constant throughout the time dynamics. The contribution from the lower orbitals is negligible. The many-body calculations are performed using $M = 12$ orbitals. The convergence of these results is detailed in the supplementary material. 

The time evolution of the occupations in the natural orbitals after the intermediate rotation quench [Fig.~\ref{fig:occupation_four}($a_2$)] exhibits a deviation from the four-fold fragmentation. The occupation of the first natural orbital, $\frac{n_1}{N}$, starts to increase from $0.25$ after  $t \ge 40$. Similarly, the occupation of the second natural orbital, $\frac{n_2}{N}$, starts rising around the same time, though at a slower rate than that of $\frac{n_1}{N}$. On the other hand, the occupations in the third natural orbital, $\frac{n_3}{N}$, and the fourth natural orbital, $\frac{n_4}{N}$, start to decrease from $0.25$, with $\frac{n_4}{N}$ declining at a faster rate than $\frac{n_3}{N}$. These trends indicate the buildup of coherence over time following the intermediate rotation quench.

Fig.~\ref{fig:occupation_four}($a_3$) presents the time evolution of the occupations of the first eight natural orbitals after the large rotation quench. As already mentioned, this calculation is performed for $N=4$ bosons and $M=28$ orbitals. The initial 25\% occupancy of the first four orbitals remains almost four-fold fragmented, though it undergoes a small fluctuation in this case. Thus, large rotational quench exhibits minor differences compared to smaller quenches.

\subsubsection{Time evolution of the average angular momentum }
The time evolution of the average angular momentum for small and intermediate quenches is presented in Fig.~\ref{fig:occupation_four}(lower row). Fig.~\ref{fig:occupation_four}($b_1$) illustrates the expectation value of the angular momentum over time for the small rotation quench. The angular momentum exhibits periodic oscillations, ranging from the maximum value of $\sim 19.49$ to the minimum value of $\sim 16.7$. The average angular momentum computed from the mean-field description is in good match with the many-body results throughout the time dynamics. For further analysis, we conducted the Fourier transformation of this data and identified one prominent frequency of the oscillations [inset of Fig.~\ref{fig:occupation_four}($b_1$)]. We have correlated the angular momentum and the density dynamics. The density dynamics exhibits a single mode, i.e., anti-clockwise rotation of the sub clouds about their potential minima. This unique motion manifests as an essentially single frequency in the Fourier spectrum.

In the intermediate quench scenario, the angular momentum oscillates over time with varying amplitudes [Fig.~\ref{fig:occupation_four}($b_2$)]. The secondary oscillation pattern shows a periodicity of $t\sim 31$. This oscillation ranges from the maximum value of $\sim 19.49$ to the minimum value of $\sim 10.31$. The mean-field calculation closely aligns with the many-body results initially (up to $t = 6 $) but differs slightly at later times. We aim to connect the behavior of the angular momentum with the density dynamics. Initially, the four density clouds rotate around their axes, resulting in oscillations with large amplitude of the angular momentum. However, at later times ($t \geq 6$), a visible vortex emerges in each sub cloud (see Fig.~\ref{fig:density_four_2.0}), resulting in a reduction of the angular momentum oscillation amplitudes. The vortex becomes more prominent around $t \sim 15$, coinciding with the lowest oscillation amplitude in the angular momentum. Subsequently, the vortices gradually fade away, restoring high-amplitude angular momentum oscillation. By $t \sim 31$, the vortices disappear completely, and the oscillation amplitude returns to the initial level. This cyclic process continues periodically until $t =200$. Thus, the emergence of one vortex in each density sub-cloud leads to the reduction in the amplitude of the angular momentum oscillations. For a deeper analysis, we performed the Fourier transformation of the many-body angular momentum data [inset of Fig.~\ref{fig:occupation_four}($b_2$)]. We find mainly two primary frequencies of these oscillations, which correspond to two rotational modes emerging during the dynamics. 
We take a closer look at the vortex dynamics and find a clear connection between how the angular momentum changes and how the vortices move. A more detailed explanation and related videos can be found in the supplementary materials.

The time evolution of the angular momentum for large rotation quench is plotted in Fig.~\ref{fig:occupation_four}($b_3$). After the quench, the average angular momentum exhibits non-periodic fluctuations during the time evolution. Moreover, the results obtained from the many-body calculations differ notably from the mean-field results. Specifically, the mean-field average angular momentum displays considerably greater amplitude fluctuations than its many-body counterpart. When comparing the density evolution (Fig.~\ref{fig:density_1.6}) with the average angular momentum, a notable distinction emerges. In the density evolution, the many-body density appears to be more diffused and uniform compared to its mean-field counterpart at the same time point. This discrepancy in density distribution is the underlying reason why the average angular momentum is higher in the mean-field results compared to the many-body results. The Fourier analysis of the mean-field and many-body angular momentum shows multiple peaks [inset of Fig.~\ref{fig:occupation_four}($b_3$)] that indicate the presence of several rotational modes during the dynamics. \\

Analyzing all the observables, it is found that introducing an additional degree of symmetry in comparison with the elongated trap geometry does not lead to any qualitatively new kinds of dynamics. Also, the dynamics observed for each rotation quench strength exhibit quantitative similarity  both from the many-body and mean-field perspectives. Still, the dynamics in the perfectly symmetric trap are completely different. Therefore, it can be concluded that if one gradually introduces more symmetry into the elongated trap, the dynamics will change because the symmetric trap is essentially an infinite-fold version of the elongated trap.

\section{Discussion and concluding remarks}\label{conclusion}
Rotation gives rise to a rich array of intriguing phenomena in ultracold systems, including the nucleation of quantized vortices, the onset of quantum turbulence, splitting of the density distribution, and so on. The present work centers on the specific case where the application of sufficiently strong rotation can lead to the splitting of the density profile~\cite{sunayana_scirep,beinke_rotation2}. However, the behavior of these split bosonic clouds in response to an abrupt change in the rotation frequency remains largely unexplored. Our study provides a comprehensive examination of the mean-field and the many-body dynamics in these trapped BECs following a rotation quench. For this purpose, we start with bosons confined in anharmonic potentials, where the initial density is split due to strong rotation. Our study is threefold: (i) rotation quench in the symmetric trap; (ii) rotation quench in the elongated trap where the symmetry of the potential breaks along x-direction; and (iii) rotation quench in the four-fold symmetric trap, reintroducing an additional degree of symmetry with respect to the elongated trap. The analysis includes measures of the one-body density, occupations of the natural orbitals, the expectation values of the angular momentum operator, and the variance of several observables.

When we suddenly introduce a rotation quench in the rotationally symmetric anharmonic trap, the distribution of the bosons (one-body density) does not change. The initial angular momentum of the system remains constant throughout the dynamics. Basically, in the symmetric trap, angular momentum is a good quantum number and prohibits any processes that could lead to the transfer or exchange of angular momentum. Hence, no change is observed either in the density or the angular momentum throughout the dynamics. The initial state is highly condensed at the many-body description and remains highly condensed throughout the dynamics. This is why the mean-field results closely match the many-body results. The monotonous dynamical behavior observed in all the measured quantities in the symmetric trap motivates us to investigate rotation quench dynamics in the asymmetric anharmonic traps, where the conservation of angular momentum is not applicable.

Post-rotation quench dynamics in the elongated trap yield more intriguing outcomes. Our analysis categorizes the quench dynamics into three regimes, characterized by the magnitude of the rotational quench: small, intermediate, and large quenches. The initial state is two-fold fragmented from the many-body perspective and fully condensed, of course, in the mean-field study. In both calculations, the initial density splits into two along the x-direction due to the initial rotation.  For the small rotation quench, the two density clouds only oscillate in an arc along the y-direction. The state remains two-fold fragmented throughout the dynamics, and the average angular momentum oscillates with a single frequency. The mean-field and the many-body results align closely. Following the intermediate rotation quench, the density dynamics display two types of motion. One type resembles that of the small rotation quench. The other type involves rotation along the axis passing through potential minima, where one visible vortex appears in each density sub-cloud and rotates along with its respective sub-cloud. Here also, the mean-field and many-body results are in close agreement. The many-body fragmented state shows a signature of buildup of coherence. The average angular momentum exhibits oscillations with two major frequencies, mirroring the two types of the density dynamics. During the large rotation quench, the density dynamics show a distinct pattern. Unlike in previous quenches, the rotational ``kick" is strong enough for the two separate density clouds to rotate in a circular ring. As they rotate, the density exhibits different shapes corresponding to different average angular momentum. Significant differences between the mean-field and many-body results are observed in both the magnitude and the timescales.

To observe how the system evolves over time when we restore some symmetry to the elongated trap, we investigate the rotation quench dynamics in the four-fold symmetric trap. In contrast to the elongated trap, here symmetry is maintained between the x- and y-directions. The initial density splits into four distinct sub clouds because of the strong rotation. We initiate the quench dynamics from a fully four-fold fragmented state and divide the quench into three cases. In the small rotation quench, the four-fold fragmented state remains four-fold fragmented throughout the time dynamics. The density of the bosons exhibits a unique motion, characterized by the rotation of each sub cloud. The angular momentum displays oscillations with a small amplitude. After the intermediate quench, the four-fold fragmented state  show a signature of buildup of coherence. The angular momentum exhibits periodic oscillations with two major frequencies, corresponding to the two rotational modes observed in the density dynamics. The first mode involves the rotation of each density sub cloud, while the second includes vortices that rotate synchronously along with the sub clouds. The behavior of  the angular momentum and its connection with the inclusion of vortices in the density dynamics have also been examined. Minor differences between the mean-field and many-body results are observed. The large rotation quench presents a more complex scenario. With numerous states participating in the dynamics, obtaining converged results becomes challenging when considering $N=8$ bosons. Therefore, the dynamics under large rotation quench is studied with the minimal system of $N=4$ bosons and exhibit, as one may anticipate, significant deviations between the mean-field condensed state and many-body fragmented state. Furthermore, no periodic pattern is observed in the time evolution of the density and the angular momentum.
In both the elongated and four-fold symmetric traps, there is no sharp boundary distinguishing the effects of different rotational quenches. The initially high rotation maintains the separation between the clouds. When the difference between the initial and final rotation frequencies is small, the density preserves the identity of the initial ground state. However, when this difference is large, the separated clouds are allowed to merge, resulting in highly complex dynamics. The strength of the quench can also be estimated by the difference in vortex numbers between the initial and final ground states. Larger differences lead to more intense vortex dynamics as the system adjusts to new stable conditions.

In the symmetric trap, rotation quenches show a consistent behavior. Conversely, in the asymmetric confinements, such as the elongated and the four-fold symmetric trapping potentials, the system exhibits rich dynamical behavior. These  range from the emergence of vortex states to changing patterns in angular momentum and various time-dependent behaviors in the mean-field and the many-body calculations. Yet, no substantial qualitative distinctions in the dynamical behavior are observed between the elongated and the four-fold symmetric potentials. However, increasing the degree of symmetry in the elongated trap and four-fold symmetric traps may lead to different dynamics, as the symmetric trap is essentially an infinite-fold version of the elongated and four-fold symmetric traps. 
Our numerical simulations reveal intricate density distributions that open new avenues for studying rotational effects in other quantum systems. As the density of a rotating BEC splits, the resulting Josephson junction dynamics arise naturally due to the rotation, eliminating the need for an external potential barrier to separate the cloud.
As an extension, rotation-mediated bosonic Josephson junctions have been studied in both position and momentum spaces~\cite{future1}. Josephson-like oscillations in supersolids reveal unique dynamics linked to their internal structure, resembling arrays of Josephson junctions~\cite{future2}. This opens another new avenues for studying quantum dynamics under rotation for supersolid systems.

\section*{Acknowledgements}
We thank Paolo Molignini for useful discussions. This work is supported by the Israel Science Foundation (ISF) grant no. 1516/19. Computation time on the High Performance Computing system Hive of the Faculty of Natural Sciences at University of Haifa and the High-Performance Computing Center Stuttgart (HLRS) is gratefully acknowledged.


\end{document}